\documentclass[]{spie}  

\usepackage{amsmath,amsfonts,amssymb}
\usepackage{graphicx}
\usepackage[colorlinks=true, allcolors=blue]{hyperref}

\usepackage{xcolor}
\usepackage{soul}

\title{Exact Solution of The Two-Reflector Optical System Meeting Herschel's Condition}

\author{Giuseppe Orlando}
\affil{Thales Alenia Space Italy, Via Saccomuro 24, 00131 Rome, Italy}

\authorinfo{Send correspondence to\\Giuseppe Orlando: giuseppe.orlando@thalesaleniaspace.com, +39 0641514334}

\begin{document} 
\maketitle

\begin{abstract}
Exact solution of stigmatic two-reflector optical system in presence of Herschel's condition is demonstrated. Details of how the solution is calculated are reported. Ray tracing verification on different optical systems validates the obtained results.  
\end{abstract}


\section{INTRODUCTION}
\label{sec:intro1}  

The design problem of a stigmatic two-reflector is a fundamental problem in optics. 
Often not only the stigmatism but also aplanatism or Herschel's condition could be important in the design procedure [\citenum{King}]. In [\citenum{ww}] has been developed a numerical method to attain this purpose. 
A more stable numerical representation of Wassermann and Wolf equations can be found in [\citenum{Yam}]. Aplanatism and Herschel's condition are known to be the two fundamental conditions for optical system design [\citenum{Born86}] and are particular cases of the generalized sine condition described in [\citenum{Elaz}].
The aplanatic optical system is free from spherical aberration and circular coma [\citenum{Born86}]. In order to be free of spherical aberration and coma, the aplanatic system shall satisfy the Abbe sine condition [\citenum{Abbe1881}]. Since the sine condition gives information about the quality of the off-axis image in terms of the properties of axial pencils it is of great importance for optical design [\citenum{Born86}]. 
Abbe's sine condition is a prerequisite for an optical system which need to image an extended flat object, e.g. for a photographic camera object, a reduction objective but also a microscope or an astronomical telescope.
The two-reflector aplanatic stigmatic system has been analytically solved by A. K. Head in [\citenum{Head2}] in a axially symmetric frame. As demonstrated in [\citenum{Head3}] for this symmetric case, only one quadrature involves the reflector surface solution. Analytical solutions for aberration-free telescopic two-reflector system  can be found in [\citenum{SchW1905}], [\citenum{Lind01}], [\citenum{Lind02}] and [\citenum{Bra}].   

The Herschel's condition, first introduced in [\citenum{Hers}], is explained in [\citenum{Born86}] and it is the complementary of Abbe sine condition in fact when satisfied, an element of the axis in the immediate neighborhood of object will be imaged sharply by a pencil of rays, irrespective of angular divergence of the pencil. Herschel's condition is requested when a system needs to operate at different magnifications, e.g. a narrow field telescope for both remote and close observations. The Abbe sine condition and the Herschel's condition cannot hold simultaneously [\citenum{Born86}].
Abbe and Herschel's conditions are the most important in optics but for some applications (e.g a fisheye lens) a different angular law describing relation between object and image space can be requested. 
Exact equations and numerical solutions for aplanatic singlet lens are reported in [\citenum{Acu1}] and in [\citenum{Acu2}] for the Herschel's condition using implicit nonlinear differential equation. A method that solves numerically an explicit ordinary differential equation for a prescribed ray mappings can be found in [\citenum{Byk}].

In this paper we will consider only axially symmetric optical systems and homogeneous and isotropic medium.  
Section 2 introduce the Abbe sine condition.
In Section 3 we present the Herschel's condition and in Section 4 we present the aplanatic two-reflector solution of [\citenum{Head2}]. Section 5 describe the equivalent Herschel's condition while in Section 6 we solve the differential equation associated to the Herschel's condition for a two-reflector. In Section 7 we simulate examples where we use the proposed solution to design a stigmatic two-reflector systems meeting the Herschel's condition. In Section 8 we finally compare the caustic of the Herschel's and the aplanatic optical systems.
We simulate optical systems by using the Geometrical Optics (GO), and Ray Tracing through software Ticratool \textregistered.

   \begin{figure} [ht]
   \begin{center}
   \begin{tabular}{c} 
   \includegraphics[height=8cm]{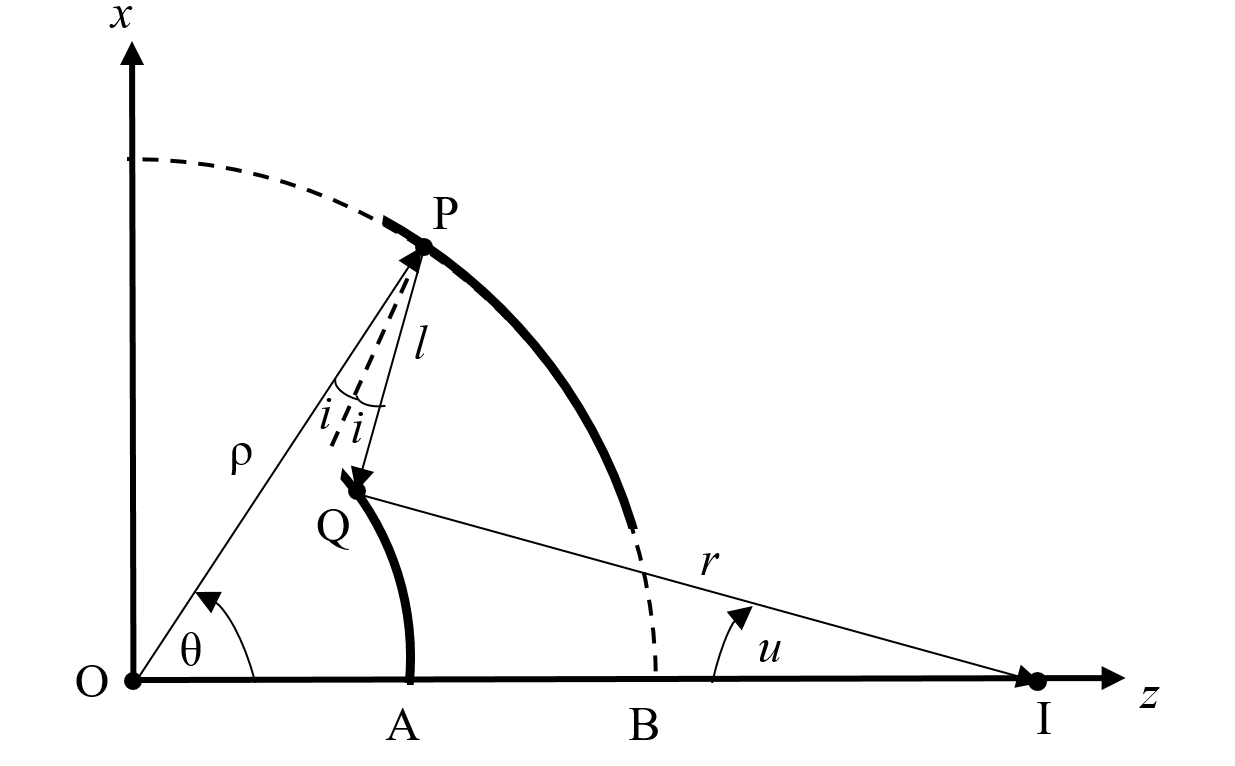}
	\end{tabular}
	\end{center}
   \caption[example] 
   { \label{fig:video-example} 
Stigmatic two-reflector optical system }
   \end{figure} 
   
\section{The Abbe Sine Condition}

The Abbe sine condition of a stigmatic optical system composed by two axially symmetric surfaces $1$ and $2$ is [\citenum{Born86}],[\citenum{Welf01}]. 
\begin{equation}
\sin \theta = m \sin u  ,
\end{equation}
where $\theta$ is the angle relative to the optical axis of any ray as it leaves the on-axis object point, $u$ is the angle of the same ray as it reaches the on-axis image point, and $m$ is the constant magnification factor of the optical system. We assume that the optical system is immersed in air. The optical system satisfying Abbe sine condition is free of spherical aberration and coma.

\section{The Herschel's condition}

The Herschel's condition of a stigmatic optical system composed by two axially symmetric surfaces $1$ and $2$ is reported in [\citenum{Born86}],[\citenum{Ste}],[\citenum{Hop}]: 
\begin{equation}
1-\cos \theta = h (1-\cos u)  ,
\end{equation}
and can be equivalently written as
\begin{equation}
\sin^2 (\theta/2) = h \sin^2 (u/2)  ,
\end{equation}
where $\theta$ is the angle relative to the optical axis of any ray as it leaves the on-axis object point, $u$ is the angle of the same ray as it reaches the on-axis image point, and $h>0$ is the Herschel's factor of the optical system. When the Herschel's condition is satisfied, an element of the axis in the immediate neighborhood of $O$ will be imaged sharply by a pencil of rays, irrespective of angular divergence of the pencil. 

\section{Exact solution of Stigmatic Two-Reflector Optical System with Abbe Sine Condition}

The two-reflector optical system with two foci, rotational symmetry and Abbe sine condition has been studied by [\citenum{Head2}]. The author solves an Ordinary Differential Equation (ODE) and we report this solution below because it is propaedeutic for our purpose.

Consider the optical system of fig. 1. From the Malus and Dupin theorem [\citenum{Born86}] we can write the Optical Path Length (OPL) as 
\begin{equation}
OP+PQ+QI=OB+BA+AI  ,
\end{equation}
that, in polar coordinates, can be written as
\begin{equation} \label{OPL}
\rho+l+r=\rho_0+l_0+r_0 ,
\end{equation}
where $\rho_0=OB$, $l_0=BA$ and $r_0=AI$ have been defined on the paraxial ray path (see fig. 1). 
The reflection laws for the first mirror can be written as [\citenum{Corn}], [\citenum{Law}]
\begin{equation} \label{d0}
\cfrac{1}{\rho}\cfrac{d\rho}{d\theta}=-\tan i  ,
\end{equation}
or equivalently
\begin{equation} \label{d1}
\cfrac{d}{d\theta} \left ( \cfrac{1}{\rho}\right )=\left (\cfrac{1}{\rho} \right )\tan i.
\end{equation}
The general expression of $\tan i$ is calculated in the Appendix A.1. By inspection of $\tan i$ we observe that it is a function of initial conditions, $\rho$, $\theta$ and $u$ through the ratio
\begin{equation} \label{t1}
\cfrac{1-\cos{\theta}}{\sin{\theta}}=\tan \cfrac{\theta}{2},
\end{equation}
and 
\begin{equation} \label{t2}
\cfrac{1-\cos{u}}{\sin{u}}=\tan \cfrac{u}{2}.
\end{equation}
Eq. \eqref{t1} is the ratio of left-side hand of Herschel and Abbe sine condition, and eq. \eqref{t2} of right-side hand of them multiplied by $h/m$. It seems that there is an intrinsic relation between the two fundamental conditions of optics, and this fact will be taken into account in the calculation of exact solution of two-reflector meeting Herschel's condition.

We have four equations to solve, two algebraic (the OPL and the Abbe condition) and two differential (the two reflection laws for the two surfaces). One of the four is redundant.
Solution related to surface 1 is summarized by following polar function [\citenum{Head2}]:

\begin{equation} \label{e1}
\begin{split}
\cfrac{l_0}{\rho}&=\cfrac{1+k}{2k} + \cfrac{1-k}{2k} \cos{\theta} +\\ &+\left(\cfrac{l_0}{\rho_0}-\cfrac{1}{k} \right)\left(\cfrac{\gamma}{1+m} \right)^{-1} 
\left( \cfrac{\gamma-1+m}{2m} \right)^{\alpha}  
\left( \cfrac{\gamma-m+1}{2} \right)^{\beta}  
\left( \cfrac{k+1}{2m+2}\gamma -\cfrac{k-1}{2} \right)^{2-\alpha-\beta}  .
\end{split}
\end{equation}
where 
\begin{equation}
\begin{cases}
k=(\rho_0+r_0)/l_0 \\
\alpha=\cfrac{mk}{mk-1} \\
\beta=\cfrac{m}{m-k}  \\
\gamma=\cos \theta+\sqrt{\strut m^2-\sin^2{\theta}} 
\end{cases}
\end{equation}
The solution for the second mirror $r$ is deduced for symmetry substituting $\theta$ with $u$, $m$ with $1/m$ and matching the initial condition $r=r_0$ for $u=0$.
Solution \eqref{e1} will be compared with the exact stigmatic solution meeting Herschel's condition. 

\section{Equivalent Herschel's condition}

Before demonstrating our main result we establish a relationship between Abbe sine condition and Herschel's condition. 
In other words we want to establish a direct law that relates the two fundamental conditions solving the following system
\begin{equation} \label{e2b}
\begin{cases}
\sin \theta = \mu \sin u \\
(1-\cos{\theta})=h(1-\cos{u})
\end{cases}
\end{equation}
where we assume $\mu=\mu(\theta)$ and $h>0$ constant. Substituting the second equation into the square of the first we get
\begin{equation} \label{e2}
\mu=\sqrt{h\cfrac{1+\cos{\theta}}{1+\cos{u}}}=h\sqrt{\cfrac{1+\cos{\theta}}{2h+\cos{\theta}-1}}.
\end{equation}
Eq. \eqref{e2} is substantially an equivalent of the Herschel's condition written through the ratio $\sin{\theta}/\sin{u}=\mu$ and it is exact, in fact is the unique solution of the system \eqref{e2b}.

\section{Exact Solution of two-reflector Meeting Herschel's Condition}
The statement of the design of Herschel's two-reflector problem is substantially the same as described in section 4, with the difference in calculating the $\tan i$. All details of this calculation are reported in Appendix A.1.
The main step of the problem is to solve eq. \eqref{d1}. Multiplying both side of eq. \eqref{d1} by $l_0$ we obtain
\begin{equation} \label{d2_0}
\cfrac{d}{d\theta}\left(\cfrac{l_0}{\rho}\right)=\left (\cfrac{l_0}{\rho} \right )\tan i,
\end{equation}
This leads to 
\begin{equation} \label{d2}
\cfrac{d}{d\theta}\left(\cfrac{l_0}{\rho}\right)=\left(\cfrac{l_0}{\rho}\right)F(\theta)+G(\theta),
\end{equation}
where 
\begin{equation} 
F(\theta)=-\cfrac{1+\cfrac{k\mu}{h}}{\cot{\cfrac{\theta}{2}}-\cfrac{k\mu}{h}\tan{\cfrac{\theta}{2}}} ,
\end{equation}
and 
\begin{equation} 
G(\theta)=\cfrac{1+\cfrac{\mu}{h}}{\cot{\cfrac{\theta}{2}}-\cfrac{k\mu}{h}\tan{\cfrac{\theta}{2}}} .
\end{equation}
In this calculation we already used eq. \eqref{h1} and eq. \eqref{e2}.
Solution of \eqref{d2} is demonstrated in the Appendix A.2. 
We report this solution for surface 1, i.e. 
\begin{equation} \label{s1}
\begin{split}
\cfrac{l_0}{\rho}&=\cfrac{1+k}{2k} + \cfrac{1-k}{2k}\cos{\theta}+ \\ &+\left ( \cfrac{l_0}{\rho_0} -\cfrac{1}{k}\right ) \left ( 1+\tan^2\cfrac{\theta}{2} \right )^{-1} \left (\cfrac{\lambda}{\sqrt{h}} - \cfrac{k}{\sqrt{h}}\tan^2\cfrac{\theta}{2}  \right ) \left ( \cfrac{2k\lambda-h+1-E}{2k\sqrt{h}-h+1-E} \right )^{\frac{h-1}{E}}\left ( \cfrac{2k\lambda-h+1+E}{2k\sqrt{h}-h+1+E} \right )^{\frac{1-h}{E}},
\end{split}
\end{equation}
where we have set 
\begin{equation}
\lambda=\sqrt{(h-1)\tan^2 \cfrac{\theta}{2}+h },
\end{equation}
and 
\begin{equation}
E = \sqrt{1+h(4k^2+h-2)}.
\end{equation}
From reversibility of the system the second mirror is obtained from the first mirror substituting  $(\rho,\theta)$ with $(r,u)$, $h$ with $1/h$ and $(\rho_0,r_0)$ with $(r_0,\rho_0)$ inside the main eq. \eqref{s1}, i.e.
\begin{equation}  \label{s2}
\begin{split}
\cfrac{l_0}{r}&=\cfrac{1+k}{2k} + \cfrac{1-k}{2k}\cos{u}+ \\ &+\left ( \cfrac{l_0}{r_0} -\cfrac{1}{k}\right ) \left ( 1+\tan^2\cfrac{u}{2} \right )^{-1} \left (\cfrac{\xi}{\sqrt{H}} - \cfrac{k}{\sqrt{H}}\tan^2\cfrac{u}{2}  \right ) \left ( \cfrac{2k\xi-H+1-S}{2k\sqrt{H}-H+1-S} \right )^{\frac{H-1}{S}}\left ( \cfrac{2k\xi-H+1+S}{2k\sqrt{H}-H+1+S} \right )^{\frac{1-H}{S}},
\end{split}
\end{equation}
where we have set $H=1/h$, with 
\begin{equation}
\xi=\sqrt{(H-1)\tan^2 \cfrac{u}{2}+H },
\end{equation}
and 
\begin{equation}
S = \sqrt{1+H(4k^2+H-2)}.
\end{equation}
We observe that these solutions are singular if we set $h=1$. But this means $1-\cos \theta = 1- \cos u$, and therefore $\theta=u$. This condition will generates polar equations of conics about a focus.
Comparing \eqref{s1} with \eqref{e1} we see that the particular solution is exactly the same and at the end is a conic function. The second part is for both a combination of exponential. To the best of our knowledge it is the first time that it is reported the exact solution of the two-reflector optical system meeting Herschel's condition.

\section{Design of Two-Reflector meeting Herschel's condition}

Differential equations of stigmatic optical system meeting Herschel's condition can be found in [\citenum{ww}][\citenum{Born86}] or in [\citenum{Yam}] with a more numerical stable frame. Up to now the only way to solve this equations is using standard numerical  routines for differential equations.
In mathematical literature these problems are identified as Semi-explicit  Algebraic Differential Equations (SADE), (see [\citenum{Asch}] for numerical way to solve this equations).
We have a new two-reflector solution therefore we can design different types of optical systems. 

   \begin{figure} [ht]
   \begin{center}
   \begin{tabular}{c} 
   \includegraphics[height=6.5cm]{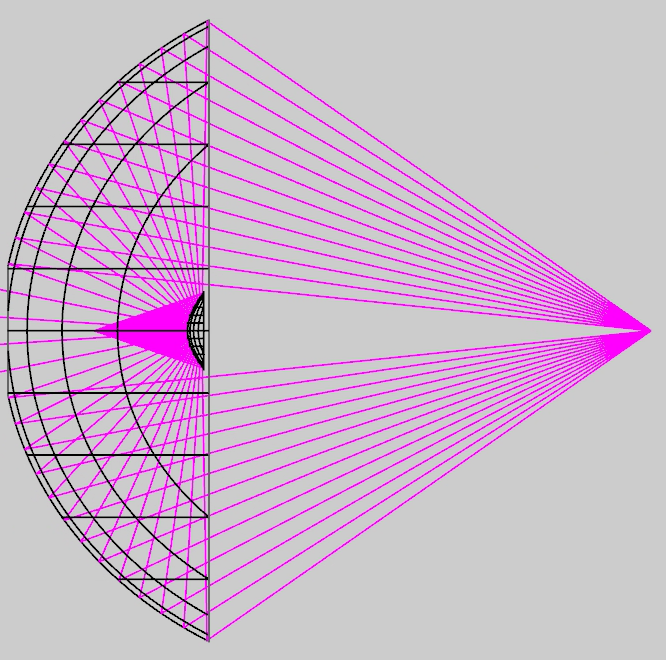}
	\end{tabular}
	\end{center}
   \caption[example] 
   { \label{fig:video-example1b} 
Herschel's optical system designed with eq. \eqref{s1}-\eqref{s2}, optical parameters are $h$=0.3, $\rho_0$=0.5m, $r_0$=3.5m, $l_0$=1m. The NOPD of \eqref{NOPD} is less than $3.5\times10^{-16}$. }   
   \end{figure} 

   \begin{figure} [ht]
   \begin{center}
   \begin{tabular}{c} 
   \includegraphics[height=7.0cm]{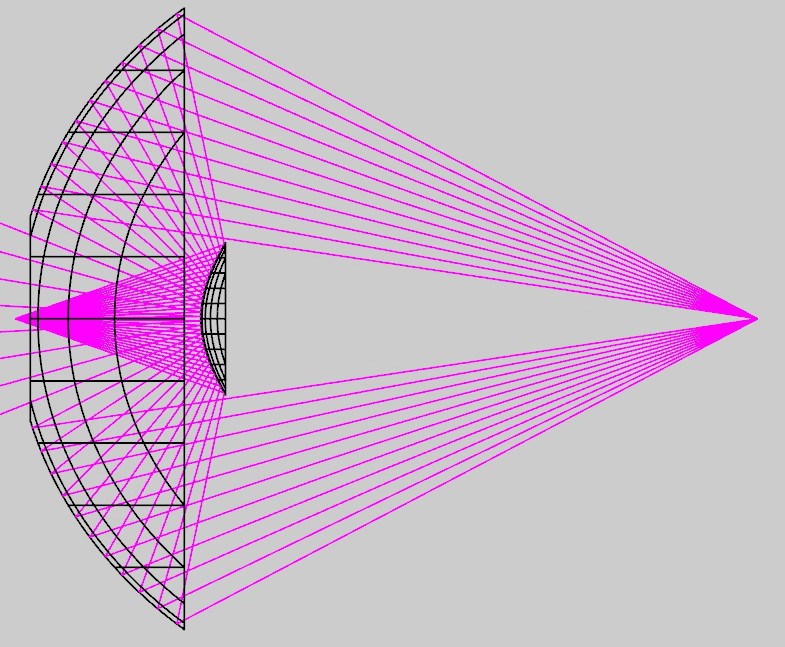}
	\end{tabular}
	\end{center}
   \caption[example2] 
   { \label{fig:video-example2} 
Herschel's optical system designed with eq. \eqref{s1}-\eqref{s2}, optical parameters are $h$=0.5, $\rho_0$=1.2m, $r_0$=4.8m, $l_0$=1.2m. The NOPD of \eqref{NOPD} is less than $5.9\times10^{-16}$. }
   \end{figure} 

   \begin{figure} [ht]
   \begin{center}
   \begin{tabular}{c} 
   \includegraphics[height=7.0cm]{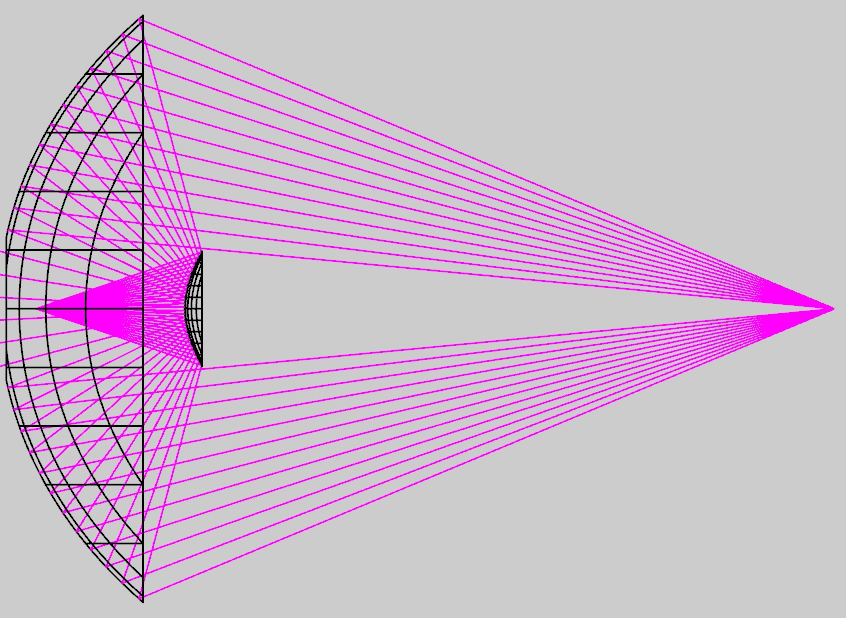}
	\end{tabular}
	\end{center}
   \caption[example3] 
   { \label{fig:video-example3} 
Herschel's optical system designed with eq. \eqref{s1}-\eqref{s2}, optical parameters are $h$=0.7, $\rho_0$=0.8m, $r_0$=4.5m, $l_0$=1m. The NOPD of \eqref{NOPD} is less than $9.9\times10^{-16}$. }
   \end{figure} 

To evaluate ray tracing performance we define the Normalized Optical Path Difference (NOPD) as
\begin{equation} \label{NOPD}
NOPD=\left|  \cfrac{LSH - RSH}{RSH} \right|,
\end{equation}
where LSH and RSH refers to left and right-side hand of eq. \eqref{OPL}.

Three examples of stigmatic two-reflector calculated by eq.\eqref{s1}-\eqref{s2} are reported in fig 2, fig.3 and fig. 4 where Herschel's factor assumes values $0.3$, $0.5$ and $0.7$ respectively. For fig. 2 the worst case NOPD is less than $3.5\times10^{-16}$, for fig. 3 is $5.9\times10^{-16}$ and for fig.4 is $9.9\times10^{-16}$. These values provide an excellent OPL satisfaction.

\section{Caustic comparison}
In this paragraph we want to compare the caustic of two systems: the Herschel's optics and the equivalent aplanatic optical system with same optical parameters (designed with \eqref{e1}). As can be seen form \eqref{e2} an Herschel's optical system with parameter $h$ can be compared with an aplanatic optical system  with parameter $m\approx\sqrt{h}$ (in fact for $\theta=0$, $\mu(0)=\sqrt{h}$ ).

For a single reflector the two caustics (sagittal and tangential) can be analytically calculated for example with [\citenum{Yamp}]. 
In case of a two reflector system the calculation is more complicated and we can use the description of [\citenum{Off}] with the help of the $k$ function. Anyway the solution is cumbersome and we don't report the result. We prefer a more pragmatic approach showing directly the caustic generated by the ray tracing simulation.  

   \begin{figure} [ht]
   \begin{center}
   \begin{tabular}{c} 
   \includegraphics[height=14.0cm]{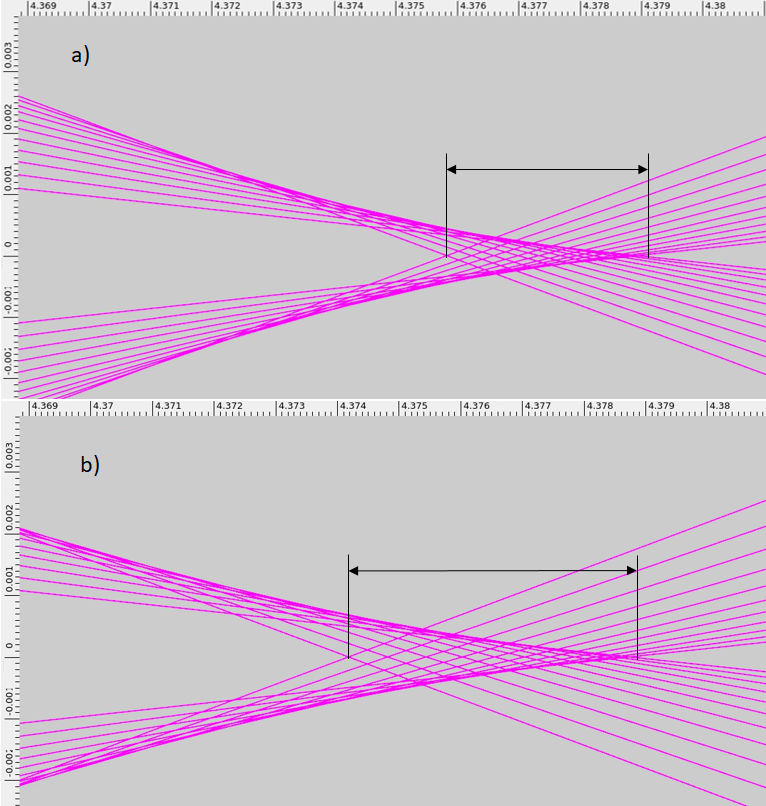}
	\end{tabular}
	\end{center}
   \caption[example4] 
   { \label{fig:video-example4} 
Caustic of the two-reflector system for an axial shift of $\Delta z=+100mm$. a) Herschel's optical system designed with eq. \eqref{s1}-\eqref{s2}, optical parameters are $h$=0.7, $\rho_0$=0.8m, $r_0$=4.5m, $l_0$=1m. b) Aplanatic optical system designed with \eqref{e1} and same optical parameters with magnification $m=\sqrt{0.7}$. The length of the sagittal caustic is highlighted in figures.  }
   \end{figure} 
From fig. 5 it is interesting to note that for the aplanatic system the sagittal caustic is increased respect the Herschel's system of about $44\%$; the tangential caustic is widely more extended for the aplanatic optical system.

\section{Conclusion}
We demonstrated the exact solution of the two-mirror optical system meeting the Herschel's condition and this follow the strictly related aplanatic solution published by A. K. Head in 1957 [\citenum{Head2}]. Despite efficient numerical methods for optical design we think that exact solutions already represents fundamental benchmarks that are useful in the field of simulation and validation problems.

\appendix    

\appendix 

\section{1}
\label{sec1:misc6}

We briefly describe main steps to calculate $\tan i$ in case of two-reflector optical system that are extracted from [\citenum{Head2}]. 
For main parameters we refer to fig. 1.
The OPL of the two-reflector is specified as
\begin{equation} \label{a0}
\rho+l+ r=\rho_0+l_0+  r_0 ,
\end{equation}
where $l$ is defined as in fig. 1. We set also
\begin{equation}
L=\rho_0-l_0+r_0 .
\end{equation}
Additional geometrical relations are
\begin{equation} \label{a1}
l \sin (2i+\theta)=\rho \sin \theta - r \sin{u} ,
\end{equation}
\begin{equation} \label{a2}
l \cos (2i+\theta)=\rho \cos \theta+r\cos{u} - L  .
\end{equation}
Multiplying \eqref{a1} by $\cos u$ and \eqref{a2} by $\sin u$ and adding we get
\begin{equation} \label{a3}
l \sin (2i+\theta+u)=\rho \sin(\theta+u)- L \sin u .
\end{equation}
Multiplying \eqref{a1} by $\sin u$ and \eqref{a2} by $\cos u$ and subtracting we get
\begin{equation} 
l \cos (2i+\theta+u)=\rho \cos(\theta+u)+r-L\cos{u}.
\end{equation}
Substituting $r$ from eq. \eqref{a0} leads to
\begin{equation} \label{a4}
l+l \cos (2i+\theta+u)=\rho \cos(\theta+u)-L\cos{u}-\rho+\rho_0+l_0+r_0.
\end{equation}
Dividing eq. \eqref{a3} by \eqref{a4} we obtain
\begin{equation} \label{a5}
\cfrac{\sin (2i+\theta+u)}{1+ \cos (2i+\theta+u)}=\cfrac{\rho \sin(\theta+u) - L \sin u}{\rho \cos(\theta+u)-L\cos{u}-\rho+\rho_0+l_0+r_0}.
\end{equation}
We can write
\begin{equation} \label{a6}
\cfrac{\sin (2i+\theta+u)}{1+ \cos (2i+\theta+u)}=\tan{\left(i+\cfrac{\theta+u}{2} \right)}=\cfrac{\tan i + \tan{\cfrac{\theta+u}{2}}}{1-\tan i \tan{\cfrac{\theta+u}{2}}}.
\end{equation}
Finally expanding eq. \eqref{a6}, using eq. \eqref{a5} and solving for $\tan i$ leads to
\begin{equation} 
\tan i=\cfrac{\rho \left( \cfrac{1-\cos{\theta}}{\sin\theta} +\cfrac{1-\cos{u}}{\sin u}\right )-l_0 \left( \cfrac{1-\cos{\theta}}{\sin\theta}\right)-(\rho_0+r_0)\left( \cfrac{1-\cos{u}}{\sin u}\right) }{l_0-(\rho+r_0)\left( \cfrac{1-\cos{\theta}}{\sin\theta}\right )\left( \cfrac{1-\cos{u}}{\sin u}\right)}.
\end{equation}
We observe that
\begin{equation} 
\cfrac{1-\cos{\theta}}{\sin{\theta}}=\tan \cfrac{\theta}{2}.
\end{equation}

In case of Herschel's condition we can use \eqref{e2}, and therefore we can write
\begin{equation} \label{h1}
\tan i=\cfrac{\cfrac{\rho}{l_0} \left( 1 +\cfrac{\mu}{h}\right )-1 -\cfrac{k\mu}{h} }{\cot{\cfrac{\theta}{2}}-\cfrac{k\mu}{h}\tan{\cfrac{\theta}{2}}},
\end{equation}
where we have used relations
\begin{equation} 
\tan \cfrac{u}{2}=\cfrac{\mu}{h}\tan \cfrac{\theta}{2},
\end{equation}
and 
\begin{equation} 
k=\cfrac{\rho_0+r_0}{l_0}.
\end{equation}

\appendix 

\section{2}
\label{sec1:misc5}

We briefly describe main steps to solve the differential equation \eqref{d2_0} in case of Herschel's condition, 
\begin{equation} 
\cfrac{d}{d\theta}\left(\cfrac{l_0}{\rho}\right)=\left (\cfrac{l_0}{\rho} \right )\tan i,
\end{equation}
where $\tan i$ is represented by eq. \eqref{h1}.
Substituting eq. \eqref{h1} into \eqref{d2_0} we obtain

\begin{equation} \label{d3}
\cfrac{d}{d\theta}\left(\cfrac{l_0}{\rho}\right)=\left(\cfrac{l_0}{\rho}\right)F(\theta)+G(\theta),
\end{equation}
where 
\begin{equation} 
F(\theta)=-\cfrac{1+\cfrac{k\mu}{h}}{\cot{\cfrac{\theta}{2}}-\cfrac{k\mu}{h}\tan{\cfrac{\theta}{2}}} ,
\end{equation}
and 
\begin{equation} 
G(\theta)=\cfrac{1+\cfrac{\mu}{h}}{\cot{\cfrac{\theta}{2}}-\cfrac{k\mu}{h}\tan{\cfrac{\theta}{2}}} .
\end{equation}
with
\begin{equation} 
\mu=h\sqrt{\cfrac{1+\cos{\theta}}{2h+\cos{\theta}-1}}.
\end{equation}

This is a first order linear ODE in $l_0/\rho$ and its solution is the sum of any particular solution of \eqref{d3} and an arbitrary multiple of the solution of the corresponding homogeneous equation
\begin{equation} \label{d4}
\cfrac{d}{d\theta}\left(\cfrac{l_0}{\rho}\right)=\left(\cfrac{l_0}{\rho}\right)F(\theta).
\end{equation}
By substituting in \eqref{d3} the trial solution $A+B\cos \theta$ with $A$ and $B$ constants, it is found that a particular solution is  
\begin{equation} \label{o1}
y_p=\cfrac{1+k}{2k} + \cfrac{1-k}{2k} \cos{\theta}.
\end{equation}
 $A+B\cos \theta$ is a particular solution if and only if the system is axially symmetric [\citenum{Head3}].
The solution of the homogeneous eq. \eqref{d4} is 
\begin{equation}
y_o=C\exp{\left (\int F(\theta) d\theta\right )},
\end{equation}
where $C$ is a constant.
As stated in [\citenum{Frol}], this integral can be evaluated in closed form only if $\tan(\theta/2)=p\tan(u/2)$ (where $p$ is a constant) and for Abbe sine condition. We add to this set the Herschel's exact solution.
In fact this integral can be evaluated with first substitution
\begin{equation}
v=\tan\cfrac{\theta}{2}
\end{equation}
leading to an integral of the type
\begin{equation}
\int \cfrac{v\left(\sqrt{(h-1)v^2+h}+k \right ) }{(v^2+1) \left(\sqrt{(h-1)v^2+h}-kv^2 \right )}dv.
\end{equation}
A second substitution $w=\sqrt{(h-1)v^2+h}$ leads to 
\begin{equation}
\int \cfrac{w(w+k)}{(w^2-1) (kw^2+(1-h)w-hk)}dw,
\end{equation}
that can be solved with partial fraction decomposition. At the end we will find logarithm and arc-tangent functions but the argument of arc-tangent is complex. Remembering that [\citenum{Geddes}],
\begin{equation}
\arctan z=\cfrac{j}{2}\log \cfrac{j+z}{j-z},
\end{equation}
with $j=\sqrt{-1}$, we come back to real logarithm function.

Collecting all substitutions and multiplicative constants the homogeneous solution is therefore
\begin{equation} \label{o2}
y_o= C\left ( \lambda^2-1 \right )^{-1} \left ( \lambda-k\tan^2\cfrac{\theta}{2} \right ) \left ( \cfrac{2k\lambda-h+1-E}{2k\lambda-h+1+E} \right )^{\frac{h-1}{E}},
\end{equation}
where we have set 
\begin{equation}
\lambda=\sqrt{(h-1)\tan^2 \cfrac{\theta}{2}+h },
\end{equation}
and 
\begin{equation}
E = \sqrt{1+h(4k^2+h-2)}.
\end{equation}
The general solution of \eqref{d3} is the sum of \eqref{o1} and \eqref{o2}, i.e. 
\begin{equation}
\cfrac{l_0}{\rho}=y_p+y_o.
\end{equation}
Finally the arbitrary constant $C$ is determined by $\rho=\rho_0$ when $\theta=0$. This gives the equation of the first mirror as
\begin{equation}
\begin{split}
\cfrac{l_0}{\rho}&=\cfrac{1+k}{2k} + \cfrac{1-k}{2k}\cos{\theta}+ \\ &+\left ( \cfrac{l_0}{\rho_0} -\cfrac{1}{k}\right ) \left ( 1+\tan^2 \cfrac{\theta}{2} \right )^{-1} \left (\cfrac{\lambda}{\sqrt{h}} - \cfrac{k}{\sqrt{h}}\tan^2\cfrac{\theta}{2}  \right ) \left ( \cfrac{2k\lambda-h+1-E}{2k\sqrt{h}-h+1-E} \right )^{\frac{h-1}{E}}\left ( \cfrac{2k\lambda-h+1+E}{2k\sqrt{h}-h+1+E} \right )^{\frac{1-h}{E}}.
\end{split}
\end{equation}

\textbf{Disclosures}. We do not have
conflicts of interest.

\textbf{Data availability}. Data underlying the results presented in this paper are
not publicly available at this time but may be obtained from the author upon
reasonable request.

\acknowledgments 
This article is dedicated to the beloved memory of my father.


\end{document}